\documentclass[12pt]{article}

\usepackage{amssymb,amsmath}
\usepackage{latexsym}
\usepackage{graphicx}
\usepackage[usenames]{color}
\usepackage{epstopdf}
\usepackage{makeidx}

\usepackage{geometry}

\usepackage{fancybox}

\usepackage{graphicx}
\usepackage{hyperref}

\usepackage{color}

\author{
  \begin{minipage}{.97\linewidth}
    \vspace{1cm}
    \begin{center}
      \begin{small}
        \textbf{F. Bourliot}\footnote{bourliot@cpht.polytechnique.fr} ${\ }^1$,
        \textbf{J. Estes}\footnote{estes@cpht.polytechnique.fr} ${\ }^2$,
        \textbf{P.M. Petropoulos}\footnote{marios@cpht.polytechnique.fr} ${\ }^1$ and
         \textbf{Ph. Spindel}\footnote{philippe.spindel@umons.ac.be} ${\ }^3$
      \end{small}
    \end{center}
    \vspace{0.5cm}
    \hspace{2cm}\begin{minipage}{.7\linewidth}
     {\it \begin{footnotesize}
  \begin{itemize}
              \item[${}^1$] Centre de Physique Th\'eorique, CNRS--UMR 7644,\\
                Ecole Polytechnique, \\
                91128 Palaiseau Cedex, France\\
            \item [${}^2$] Laboratoire de Physique Th\'eorique, CNRS--UMR 8549,\\
                Ecole Normale Sup\'erieure, \\
                24 rue Lhomond, F--75231 Paris cedex 05, France\\
            \item[${}^3$] Service de M\'ecanique et Gravitation, \\ Universit\'e de Mons, \\
                    20 Place du Parc, 7000 Mons, Belgique\\
        \end{itemize}
     \end{footnotesize}}
    \end{minipage}
    \vspace{0.5cm}
  \end{minipage}
}

\date{\today}

\title{\vspace{2cm}
 \boldmath \begin{huge}
    \textbf{$G3$-homogeneous gravitational instantons}
  \end{huge} \unboldmath
}

\newcommand{\be}{\begin{equation}}
\newcommand{\ee}{\end{equation}}

\newcommand{\gw}{\omega}
\newcommand{\gvp}{\varpi}
\newcommand{\gO}{\Omega}
\newcommand{\gL}{\Lambda}
\newcommand{\gl}{\lambda}
\newcommand{\ga}{\alpha}

\newcommand{\fkg}{\mathfrak{g}}
\newcommand{\mbR}{\mathbb{R}}
\newcommand{\bea}{\begin{eqnarray}}
\newcommand{\eea}{\end{eqnarray}}
\newcommand{\tmts}{\big( t-t_* \big)}
\newcommand{\twotmts}{\big(2 (t-t_*) \big)}

\begin{document}
\maketitle

\thispagestyle{empty}

\vspace{-15cm}
\begin{flushright}
    CPHT-RR125.1109, LPTENS--09/04
\end{flushright}

\vspace{13cm}

\begin{abstract}
We provide an exhaustive classification of self-dual four-dimensional gravitational instantons foliated with three-dimensional homogeneous spaces, \textit{i.e.} homogeneous self-dual metrics on four-dimensional Euclidean spaces admitting a Bianchi simply transitive isometry group. The classification pattern is based on the algebra homomorphisms relating the Bianchi group and the duality group $SO(3)$. New and general solutions are found for Bianchi III.
\end{abstract}

\section{Introduction}
Classical solutions of general relativity in diverse dimensions have been analyzed over the years from various perspectives. Instanton-like configurations have in particular attracted attention because of their potential role in the determination of transition amplitudes in quantum gravity. Similarly, their real-time counterpart turns out to be useful in the Hartle--Hawking formulation of quantum cosmology, even though general relativity would ultimately require a ultra-violet completion, possibly provided by strings. In fact, several classical solutions turn out to be embeddable in string theory, sometimes even as exact backgrounds, and instantons have certainly played a major role in addressing many issues, among which are those related to (super)symmetry breaking.\\

Following the paradigm of self-dual Yang--Mills instantons \cite{Belavin:1975fg}, self-duality has been successfully implemented in four-dimensional general relativity. In order to be operational, self-duality must be accompanied by some specific ansatz for the geometry $\mathcal{M}$. The usual ansatz is to assume  $\mathcal{M}$  topologically $\mathbb{R}\times \Sigma_3$ and, further, the leaves $\Sigma_3$ to be homogeneous spaces, admitting at least a three-dimensional group of motions $G3$. \\

The first solution obtained according to the above pattern is due to Eguchi and Hanson \cite{Eguchi:1978xp,Eguchi:1978gw} under the assumption of Bianchi IX geometry \emph{i.e.} with $SU(2)$ isometry. This gravitational instanton was an alternative to the earlier Taub--NUT metric, constructed from a different perspective though  \cite{Taub-nut}. Both were self-dual and $SU(2)$-homogeneous, with isometry enhancement to $SU(2) \times U(1)$. General extensions of Eguchi--Hanson solution to strict $SU(2)$ were obtained soon after \cite{Belinsky:1978ue}, observing that the equations were a special case of the Euler top -- with some peculiar inertia momenta -- known as the Lagrange system  \cite{Gibbons:1979xn}. A similar achievement for the Taub-NUT instanton revealed far more involved, and a particular solution was finally obtained in Ref. \cite{Atiyah:1978wi}. The difficulty to solve in full generality the corresponding equations was later understood in terms of non-algebraic integrability properties, as it was realized
\cite{ Takhtajan:1992qb} that the system at hand had already been set by Darboux  \cite{Darboux}, and solved extensively by Halphen \cite{halph1, halph2}, more than a century before, in terms of modular forms\footnote{A translation of the general Halphen--Darboux solutions in terms of gravitational instantons can be found in  \cite{Petropoulos:2008}. They are all plagued with naked singularities, except, marginally, for the particular solution of Atiyah--Hitchin \cite{Atiyah:1978wi}.}.\\

The above account for the Bianchi IX group raises immediately two questions: do other Bianchi groups possess similar solutions and what is the classification principle behind the appearance of distinguished classes of equations such as Lagrange versus Darboux--Halphen systems? Despite the large amount of information accumulated so far (see e.g.   \cite{ Lorenz:1983,Lorenz:1989}) and the physical interest of some -- even simple -- solutions like Kasner's
Bianchi I \cite{Kasner:1921zz}, a precise and definite answer to these questions was still missing. The aim of the present work is to tame the plethora of scattered results under a simple classification principle.
Our analysis is performed along the lines announced in \cite{Bourliot:2009fr}. It is general and exhaustive, and makes no assumptions on the geometry, other than those already quoted above. In particular, the issue of the choice of a diagonal versus non-diagonal metric in a given $G3$-invariant  frame is treated with care, as opposed to some former, more cavalier approaches for non-unimodular Bianchi groups.  As a bonus, this enables us to discover a new solution for Bianchi III, completing thereby the existing landscape.\\

Let us summarize the method and the results. Self-dual vacuum solutions satisfy $\Omega=\tilde \Omega$ where $\Omega$ is the Riemann curvature two-form\footnote{Self-duality can be imposed alternatively on the Weyl tensor, and leads thus to solutions with cosmological constant like Fubini--Study or Pedersen \cite{Pedersen:1986}. Note also that anti-self-solutions are obtained by parity or time reversal. }.  These are second-order equations and are equivalent to the first-order set obtained with the connection one-form: $\omega = \tilde \omega + A$. The one-form $A$ appears as a ``constant of motion'' of the second-order system. It stands as the anti-self-dual $\frak{so}(3)$ part of the Levi--Civita connection and must be flat.  The program is thus cast as follows: (\romannumeral1) find all possible flat $\frak{so}(3)$ connections over $G3$, and (\romannumeral2) for each of them, write the corresponding first-order equations,  and find the most general solution. The latter can represent a \emph{bona fide} geometry, but in most cases it is spoiled by naked singularities, if not everywhere degenerate like in all non-unimodular groups, except for Bianchi III. \\

In Sec. \ref{not} we provide some technical tools, useful to set our philosophy for the subsequent developments. Section \ref{sd} contains the core of the classification principle (point (\romannumeral1) above), whereas the exhaustive solution search (point (\romannumeral2)) is presented in Sec. \ref{allsol}. A last section (\ref{sdcon}) is devoted to the subtle issue of rotating (locally) the frame where $\omega = \tilde \omega + A$ into a frame where the connection is genuinely self-dual. Conclusions follow and a summary of all $G3$-invariant metrics finally available is presented in the appendix.\\

\section{Notations and general considerations}\label{not}

Inspired by applications to homogeneous cosmology, we consider Euclidean spaces admitting a $G3$ group of motion acting transitively\footnote{The three-dimensional group $G3$ acts \emph{simply} transitively on the leaves,  endowed thus with the structure of a group manifold.  Hence we exclude $H_3, H_2\times S^1$ or $S^2\times S^1$, which are the alternatives to the nine Bianchi classes.} on 3-dimensional invariant subspaces of the form ${\cal M}={\mathbb R}\times \Sigma^3$. The metric on these spaces can always be locally written as (see for instance Ref. \cite{GBS}, page 516, for a proof):
\begin{equation}{\label{genmet}}
ds^2=N(t)^2dt^2+\delta_{ab}\Theta^a\Theta^b.
\end{equation}
Let us introduce the co-frame defined by $\Theta^0=N(t)\,dt$ and $\Theta^a=\theta^a_\alpha(t)\,\sigma^\alpha$, where $\sigma^\alpha$ are invariant $1-$forms:
\begin{equation}{\label{invform}}
d\sigma^\alpha=\frac 12 c^\alpha_{\beta\gamma}\,\sigma^\beta\wedge\sigma^\gamma .
\end{equation}
From the latter we obtain
\begin{eqnarray}\label{dTh}
d\Theta^a&=&\dot \theta^a_\alpha\,dt\wedge \sigma^\alpha+\frac 12 \theta^a_\alpha\,c^\alpha_{\beta\gamma}\,\sigma^\beta\wedge\sigma^\gamma\\&=&\frac 1N\dot\theta^a_\alpha\,\theta^\alpha_b\,\Theta^0\wedge\Theta^b+\frac 12 \theta^a_\alpha\,c^\alpha_{\ \beta\gamma}\,\theta^\beta_b\,\theta^\gamma_c\,\Theta^b\wedge\Theta^c
\end{eqnarray}
providing the non vanishing structure coefficients:
\begin{eqnarray}
\beta^a_{\ 0b}&=&\frac 1N \dot\theta^a_\alpha\theta^\alpha_b\label{bab0}=-\beta^a_{\ b0}\label{beta0b},\\
\beta^a_{\ bc}&=&\theta^a_\alpha\,c^\alpha_{\ \beta\gamma}\,\theta^\beta_b\,\theta^\gamma_c=-\beta^a_{\ cb}\label{betabc},
\end{eqnarray}
and connection coefficients
\begin{eqnarray}
\gamma_{0ab}&=&\frac 12(\beta_{ab0}+\beta_{ba0})=\gamma_{0ba}\label{gam0ab},\\
\gamma_{ab0}&=&\frac 12(\beta_{ab0}-\beta_{ba0})=-\gamma_{ba0}\label{gamab0},\\
\gamma_{abc}&=&\frac 12(\beta_{abc}+\beta_{bca}-\beta_{cab})=-\gamma_{bac}.\label{gabc}\label{gamabc}
\end{eqnarray}
With respect to the basis    $\{dt,\sigma^\alpha\}$ the metric components are $g_{00}=N(t)^2$, $g_{0\alpha}=0$ and $g_{\alpha\beta}=\delta_{ab}\theta^a_\alpha\theta^b_\beta$. The Levi-Civita connection one-forms are:
\begin{eqnarray*}
&&\gw_{0i}=\gvp_{0i\alpha}\,\sigma^\alpha,\\
&&\gw_{ij}=\gvp_{ij0}\,dt+\gvp_{ij\alpha}\,\sigma^\alpha.
\end{eqnarray*}
Finally, we need the expression of the curvature components, which are:
\begin{eqnarray}
\gO_{0i}&=&d\gw_{0i}+\gw_{0k}\wedge\gw^k_{\ i}\nonumber\\
&=&\dot\gvp_{0i\alpha}\,dt\wedge\sigma^\alpha+\frac 12 \gvp_{0i\alpha}\,c^\alpha_{\ \beta\gamma}\,\sigma^\beta\wedge\sigma^\gamma\nonumber\\
&&-\gvp_{0k\alpha}\gvp^k_{\ i0}\,dt\wedge \sigma^\alpha+\gvp_{0k\beta}\gvp^k_{\ i\gamma}\,\sigma^\beta\wedge\sigma^\gamma,\\
\gO_{jk}&=&d\gw_{jk}+\gw_{j0}\wedge \gw^0_{\ k}+\gw_{jl}\wedge \gw^l_{\ k}\nonumber\\
&=&\dot\gvp_{jk\alpha}\,dt\wedge\sigma^\alpha+\frac 12 \gvp_{jk\alpha}\,c^\alpha_{\ \beta\gamma}\,\sigma^\beta\wedge\sigma^\gamma+\gvp_{j0\alpha}\gvp^0_{\ k\beta}\sigma^\alpha\wedge \sigma^\beta
\nonumber\\
&&+\gvp_{jl0}\gvp^l_{\ k\alpha}\,dt\wedge\sigma^\alpha-\gvp_{jl\alpha}\gvp^l_{\ k0}\,dt\wedge\sigma^\alpha+\gvp_{jl\alpha}\gvp^l_{\ k\beta}\,\sigma^\alpha\wedge\sigma^\beta.
\end{eqnarray}

With respect to this basis, the anti-self dual components of the curvature 2-form read as:
\begin{eqnarray}
&&\gO_{0i}-\frac 12 \epsilon_{ijk}\gO^{jk}=\nonumber\\
&&\left[\dot \gvp_{0i\alpha}-\gvp_{0k\alpha}\gvp^k_{\ i0}-\frac 12 \epsilon_i^{\ jk}\left(\dot\gvp_{jk\alpha}+\gvp_{jl0}\gvp^l_{\ k\alpha}-\gvp_{jl\alpha}\gvp^l_{\ k0}\right)\right] dt\wedge\sigma^\alpha\label {ASD0a}\\
&&+\frac12\left[\gvp_{0i\alpha}c^\alpha_{\ \beta\gamma}+\gvp_{0k\beta}\gvp^k_{\ i\gamma}-\gvp_{0k\gamma}\gvp^k_{\ i\beta}
-\epsilon_i^{\ jk}\left(\frac 12 \gvp_{jk\alpha}c^\alpha_{\ \beta\gamma}+\gvp_{j0\beta}\gvp^0_{\ k\gamma}+\gvp_{jl\beta}\gvp^l_{\ k\gamma}\right)\right]\sigma^\beta\wedge\sigma^\gamma\nonumber
\end{eqnarray}
or, introducing
\begin{equation}\label{Ip}
\bar I_{i\alpha}:=\gvp_{0i\alpha}-\frac 12 \epsilon_i^{\ jk} \gvp_{jk\alpha},
\end{equation}
\begin{equation}\label{ASDI}
\gO_{0i}-\frac 12 \epsilon_{ijk}\gO^{jk}=
[\dot {\bar I}_{i\alpha}-\bar I_{l\alpha}\gvp^l_{\ i0}]dt\wedge\sigma^\alpha+\frac 12 [\bar I_{i\alpha}c^\alpha_{\ \beta\gamma}+\epsilon_i^{jk}\bar I_{j\beta}\bar I_{k\gamma}]\sigma^\beta\wedge\sigma^\gamma.\
\end{equation}
Let us emphasize however that the one-forms $\theta^\alpha_k$ are not completely fixed, but defined up to
time-dependent $O(3)$ transformations and time-independent $GL(3,{\mathbb R})$ transformations, we have indeed not yet fixed the basis of invariant forms.

\section{Self-duality equations}\label{sd}

Self-duality requires that
\be
\gO_{0i}-\frac 12 \epsilon_{ijk}\gO^{jk}=0.
\ee
We see that to trivially obtain first integrals from this equation a sufficient (and necessary) condition is to require that
\begin{equation}\label{eqsym}\gvp_{ik0}=0\Leftrightarrow \dot\theta_{i\alpha} \theta^\alpha_k -\dot\theta_{k\alpha}\theta^{\alpha }_i=0\Leftrightarrow\dot\theta_{i\alpha}=-g_{\alpha\beta}\dot\theta^\beta_i,\end{equation}
after which (\ref{ASDI}) becomes equivalent to
\be
\dot {\bar I}_{i\alpha} = 0 \qquad \qquad \bar I_{i\alpha}c^\alpha_{\ \beta\gamma}+\epsilon_i^{jk}\bar I_{j\beta}\bar I_{k\gamma} = 0
\ee
The gauge freedom allows us to always satisfy this condition. Indeed if we have a solution  $\theta_{a\alpha}$, using a time dependent rotation we obtain $\tilde{\theta}_{a\alpha}:={\cal O}_a^{\ b}(t)\theta_{b\alpha}$ which will fulfill the symmetry condition if
\begin{equation}
 \dot {\cal O}_a^{\ b}=\frac 12 {\cal O}_a^{\ c}(\theta^\alpha_c\dot\theta_{\alpha}^b-\dot\theta_{c\alpha}\theta^{\alpha b}).
\end{equation}
This is a kinematic problem aiming to determine a sequence of rotations knowing the angular velocity so that it always admits a solution. Then the only transformations still possible correspond to the choice of the initial conditions of the kinematical problem, denoted from now on by the matrix $O_{a}^b$, and of arbitrary $GL(3,\mathbb R)$ time independent transformations, denoted from now on by the matrix $\gL^\beta_\alpha$,
\begin{equation}\label{tred}
\theta_{a\alpha}\mapsto O_{a}^{\ b}\theta_{b\beta}\gL^\beta_\alpha\ .
\end{equation}
In the gauge defined by (\ref{eqsym}), the first integrals $( \bar I_{i\alpha}=\mbox{const.})$ furnish the first order equations we need to solve:
\begin{equation}\label{Ifl}
\frac 1N\dot\theta_{i\alpha}-\theta^{-1}\,\theta_{i\gamma}\left[\left(n^{\gamma\mu}-a_\rho\epsilon^{\rho\gamma\mu}\right)g_{\mu\alpha}-\frac 12 \delta^\gamma_\alpha \, n^{\mu}_{\mu}\right]:=-\bar I_{i\alpha}\end{equation}
where
$\theta=\sqrt{\det (g_{\ \alpha\beta})}$ and the symmetric tensor density $n^{\gamma\mu}=n^{(\gamma\mu)}$ and vector $a_\rho$ are defined by the relations
\be
c^\alpha_{\ \beta\gamma}\epsilon^{\beta\gamma\mu}= 2(n^{\alpha\mu}+\epsilon^{\alpha \mu\beta}a_\beta).
\ee
In the following we shall call  the $t $ coordinate  {\it time} and speak about {\it evolution}, to describe this flow, though  the framework is Euclidean.\\

Using time re-parameterizations, we may choose a specific $N$ without loss of generality.
Equations (\ref{Ifl}) strongly suggest to adopt, in a first step at least, the gauge $N=\theta$.
Let us notice that in order to preserve this gauge condition when we make a transformation (\ref{tred}) we have to simultaneously rescale $t$ by a factor $1/\det (\Lambda^\beta_\alpha)$; then the time variable may still only be changed by a shift $t\mapsto t+t_0$. But before trying to integrate   equations (\ref{Ifl}) it is mandatory to solve the constraint equations:
\begin{equation}\label{mainalg}
\bar I_{i\alpha}c^\alpha_{\ \beta\gamma}+\epsilon_i^{jk}\bar I_{j\beta}\bar I_{k\gamma}=0\ .
\end{equation}
The solutions of these equations describe homomorphisms from the Lie algebra of the homogeneity group   $\fkg_3$ into the Lie algebra $\frak {so}(3)$:
\be
\bar I \ : \fkg_3 \rightarrow \frak {so}(3).
\ee
Depending on the subalgebra of $\frak {so}(3)$ on which we project $\fkg_3$ different simplifications can occur. Let us remind that
the only subalgebras of  $\frak {so}(3)$ are the trivial ones: $\frak {so}(3)$ and the null one $\{\bf 0\}$ and  the one-dimensional ones : $\mathbb R$ (all equivalent, {\it i.e.} linked by internal conjugation). Thus if the constants of motion  $\bar I_{i\alpha}$ are not identically zero (in which case we have the trivial homomorphism mapping $\fkg_3$ onto $\{\bf 0\}$) either $\fkg_3=\frak {so}(3)$ or the algebra $\fkg_3$ has a 2 dimensional ideal.\\

These remarks lead to the following Bianchi types to be considered, according to the rank of the matrix
$\bar I_{i\alpha}$:
\begin{itemize}
\item {\bf rank 3\,}(maximal) type IX ,
\item {\bf rank 2 }  impossible,
\item {\bf rank 1 }  types I, II, III, IV, V,VI,VII,
\item {\bf rank 0 } all Bianchi types .
\end{itemize}
But it remains to examine if all these cases can effectively be obtained \textit{i.e.} if there are no obstructions coming from (\ref{eqsym}). Let us notice that using (\ref{Ifl}), condition (\ref{eqsym}) can be written as
\begin{equation}\label{sym2}
\bar I_{i\alpha}\theta_ j^\alpha-\bar I_{j\alpha}\theta_ i^\alpha+2\,a_\rho\,\theta^{\rho k}\epsilon_{kij}=0\ .
\end{equation}
From this equation we see that for rank 0, we must have $a_\rho = 0$, which is the statement that rank 0 solutions must be from Bianchi class A, {\it i.e.} types I, II, VI$_{0}$, VII$_{0}$, VIII and IX, see table \ref{canon} for our conventions.  For the rank 1 cases, as shown in the next section, the only solutions whose metric determinant does not everywhere vanish are those of Bianchi class A and type III.\footnote{To see this, first decompose $\bar I_{i\alpha}$ as $\bar I_{i\alpha}=\lambda_i I_\alpha$ where $I_\alpha$ and $\lambda_ i$ are two triples.  For all types of Bianchi class B, except type III, the only solution to the constraint (\ref{mainalg}) is to take $I_\alpha$ proportional to $a_{\alpha}$.  In this case, one may show that (\ref{sym2}) requires the determinant of $\theta_ i^\alpha$ to vanish.}  Thus if we also require the metric determinant to not everywhere vanish, the above list is reduced to
\begin{itemize}
\item {\bf rank 3\,}(maximal) type IX ,
\item {\bf rank 2 }  impossible,
\item {\bf rank 1 }  types I, II, III, VI$_0$,VII$_0$ ,
\item {\bf rank 0 } types I, II, VI$_0$, VII$_0$, VIII, XI.
\end{itemize}
Thus, according to the rank of the mapping $\bar I$ we obtain the classification of solutions in the next section.

\section{All self-dual solutions} \label{allsol}

In this section, we obtain all the self-dual solutions categorized by the rank of $\bar I$. Some of them have been discussed in the past in \cite{Lorenz:1983,Lorenz:1989}\footnote{However in Ref. \cite{Lorenz:1983,Lorenz:1989} the metrics were assumed to be {\it a priori} diagonal, a technical assumption just introduced to facilitate the integration of the flow equations. Here we justify that they furnish the most general ones in case of rank 0 cases, and when necessary shall discuss the integration of the relevant non-diagonal metrics.}. On a similar ground of simplicity, the only rank 3 solution coming from the type IX Bianchi algebra has also been studied in \cite{Lorenz:1983,Lorenz:1989} and various other places in the past. We finally deal with the new cases, solutions of rank 1. Thanks to residual symmetries and constraints coming from the self-duality requirement, we simplify as most as possible the equations and present the most general solutions as well as some particular simplified solutions when available.
\paragraph{Rank zero} All $\bar I_{k\alpha}$ are zero. The Bianchi type must be of class $A$, ($a_\rho=0$), {\it i.e.} types I, II, VI$_{0}$, VII$_{0}$, VIII and IX. Equations (\ref{Ifl}) imply that the symmetry condition is always satisfied :
\begin{equation}\label{vsym}
\dot \theta_{i\,\alpha}\,\theta_j^\alpha=\theta_{i\,\gamma}\left(n^{\gamma\mu}-\frac 12n^\nu_\nu\,g^{\gamma\mu}\right)\theta_{j\,\mu}=\dot \theta_{j\,\alpha}\,\theta_i^\alpha\ .
\end{equation}
At an initial time, one may simultaneously diagonalize  $g_{\mu\nu}$ and put $n^{\mu\nu}$ in its canonical form (see table \ref{canon}). Thus at initial time the co-frame components $\theta_{i\,\gamma}$ may be chosen diagonal, and they remain so during their evolution. The explicit integration of the evolution equations has been described a long time ago in Ref. \cite{Lorenz:1983,Lorenz:1989}, here this class of self-dual spaces is presented in appendix A for completeness.
\begin{table}[h]
\caption{Canonical structure constants for the different Bianchi groups}
\begin{center}
\begin{tabular}
{|c|c|c|c|c|c|}\hline
  Type& a  &$n^1$&$n^2$&$n^3$&Usual\ name   \\
  \hline
  \multicolumn{6}{|c|}{Class   A} \\
  \hline
  I&   $0$ & $0$ &$0$ &$0$ & Translations \\
    II&  $0$  &$1$  &$0$ &$ 0$&Galilean  \\
    VII$_{0}$& $0$   &$1$  &$1$ &$ 0$&Euclidean  \\
    VI$_{0}$&$0$   &$1$  & $-1$&$0$ &Poincar\'e  \\
    IX&  $0$&  $1$&$1$  &$1$ & Rotation  \\
    VIII& $0$   &$1$  &$1$ &$-1$ &Lorentz  \\
    \hline
   \multicolumn{6}{|c|}{Class   B} \\
  \hline
      V& $1$  & $0$ &$0$ &$0$ &  \\
      IV  &  $1$  &$1$  &$0$ &$0$ &  \\
       VII$_h$&$h>0$    &$1$  &$1$ &$0$ &  \\
       VI$_{h\neq 1}$&$h>0$    &$1$  &$-1$ &$0$ &  \\
       III $\equiv$ VI$_1$&1   &$1$  &$-1$ &$0$ &  \\
       \hline
\end{tabular}
\end{center}
\label{canon}
\end{table}

\paragraph{Rank three}
 Of course only Bianchi IX is possible for this case of maximal rank. In an appropriate frame the matrix $n$ is the identity matrix. The self-duality condition (\ref{mainalg}) can be reinterpreted as a relation between three 3-dimensional  Euclidean vectors :
\begin{equation}\label{vecB9}
\vec I_1=\vec I_2\times \vec I_3
\end{equation}
and two similar relations obtained by circular permutation of the indices $1,2,3$. Thus these three vectors are perpendicular to each other and actually define an orthonormal frame of $E^3$. Acting with an appropriate element of $O(3)$, on the flat indices $i$, we may assume that $-(\bar I_{i\alpha})$ is the identity matrix. The integration of (\ref{Ifl}), under these assumptions, was first  discussed in \cite{Gibbons:1979xn,Darboux,halph1, halph2} and the solution is given in appendix A.

 \paragraph{Rank one}
In that case there exists two triplets $I_\alpha$ and $\lambda_ i$ such that  $\bar I_{i\alpha}=\lambda_i I_\alpha$. Condition (\ref{mainalg}) reduces to
\begin{equation}\label{R1I}
I_\alpha c^\alpha_{\beta\gamma}=0.
\end{equation}
\subparagraph{Bianchi  class A} In this case, condition (\ref{R1I}) is equivalent to
\begin{equation}\label{R1IA}
I_\alpha n^{\alpha\beta}=0,
\end{equation}
so the matrix $(n^{\alpha\beta})$ has to be singular. Consequently the Bianchi type has to be I, II, VI$_{0}$ or VII$_{0}$. The symmetry condition (\ref{sym2}) becomes then:
\begin{equation}\label{1Isym}
\lambda_i\,\rho_j-\lambda_j\,\rho_i=0\qquad \text{with} \qquad \rho_k=\theta_k^\alpha I_\alpha.
\end{equation}
It is easy to see that if the initial conditions satisfy this condition, then the equation will automatically be preserved throughout the evolution.  This can be seen by computing the evolution of $\rho_i$ using (\ref{Ifl}) to obtain:
\begin{equation}\label{DIA}
\dot \rho_i=(\frac 12 n^{\mu\nu}g_{\mu\nu}+\lambda^k\rho_k)\rho_i.
\end{equation}
Thus we find no restrictions on the existence of solutions to our self-dual problem.\\

In order to find the rank 1 solutions for Bianchi class  A, let us start with the most general expression of the matrix of  frame components :
\begin{equation}\label{gencf}
(\theta^\alpha_i)=
\left(
\begin{array}{ccc}
a  & p  &q   \\
r  & b  &  s \\
 u & v  & c
\end{array}
\right).
\end{equation}
It can be simplified by making use of the transformations (\ref{tred}). Using an $O(3)$ transformation we may assume $\lambda=(0,0,1)$ and, without altering the canonical value of the structure constants, using $GL(3,\mathbb R)$ transformations set
\begin{itemize}
\item $I_\alpha=(0,0,1)$  for Bianchi I,II
\item $I_\alpha=(0,0,Z)$  for Bianchi VI$_0$,VII$_0$
\end{itemize}
Moreover, for Bianchi type I, II, VI$_{0}$ and VII$_{0}$ (Bianchi type of class A), without loss of generality, we may also assume  the matrix (\ref{gencf}) diagonal. This can be done using the transformations (\ref{tred}), which now leave $I_\alpha$ and $\lambda_i$ invariant as well as the canonical values of the structure constants.
The integration of the resulting self-duality equations was given in ref \cite{Lorenz:1983,Lorenz:1989}.\\

For illustrative purpose let us discuss the Bianchi type VI$_{0}$. The integration proceeds as follows. First we observe that (\ref{1Isym}) imposes $u=v=0$. Equations (\ref{DIA}) then insure that if $u,v$ are chosen zero at the initial time, they will remain null. Then we remark that the evolution equations imply that the functions $q$ and $s$ are proportional to $c$:
\be
q(t)=Q\,c(t),\ s(t)=S\,c(t)
\ee
and that the transformations (\ref{tred}) of the frame that preserves the canonical choice of the structure constants, of $I_\alpha$ and $\lambda$ are given by:
\begin{equation}\label{OLVI}
O=
\left(
\begin{array}{ccc}
\pm \cos(\phi)  & \sin(\phi)  & 0  \\
\mp\sin(\phi)  & \cos(\phi)  &0   \\
0  & 0  &  1
\end{array}
\right)\qquad\text{and}\qquad
\Lambda^{-1}=\left(
\begin{array}{ccc}
\alpha  & \beta  &\gamma  \\
\beta  &\alpha   &\delta   \\
 0 & 0  & 1
\end{array}
\right)\  .
\end{equation}
Transformations (\ref{OLVI}) allow to simplify the frame components, without loss of generality. First we may put  $Q=S=0$ {\it i.e.} $q(t)=s(t)=0$ by performing a transformation with $O=Id$ and $\alpha=1$, $\beta=0$, $\gamma=- Q $ and  $\delta =- S$. Then we still have the freedom to make transformations (\ref{OLVI}) with $ \gamma=\delta=0$. Performing, at an arbitrary time $t_0$,  a new transformation with still $O=Id$ and $\alpha=\rho\,\cos(\varphi)$, $ \beta=\rho\,\sin(\varphi)$ where $\rho >0$ is arbitrary and $\varphi$ determined by\footnote{We use the subscript $0$ to indicate the value of the corresponding function at the instant considered : $a(t_0)=a_0$, etc.}
\begin{equation}\label{vphi}
\sin (2\varphi)=-\frac{2(a_0\,r_0+p_0\,b_0)}{(a_0^2+b_0^2+p_0^2+r_0^2)}
\end{equation}
we obtain a new frame such that now $a_0\,r_0+b_0\,p_0=0$. Let us notice that (\ref{vphi}) makes sense because from $(a_0\pm r_0)^2+(b_0\pm p_0)^2>0$ we deduce that its right-hand side member is always between $-1$ and $+1$. Then choosing $\Lambda=Id$ and $O$ with $\phi$ such that
\begin{equation}\label{vphi2}
\sin (\phi)\,p_0+\sin (\phi)\,b_0=0=-\sin (\phi)\,a_0+\cos (\phi)\,r_0
\end{equation}
we have obtained a diagonal frame at time $t_0$.\\

It remains then five functions to be determined: $a$, $b$, $r$, $p$ and $c$. These functions have to satisfy the equations:
\begin{eqnarray}
\dot a& =&\frac {a\, \Delta - 2 \kappa \,b}{2\,\kappa^2},\\
\dot b& =&\frac {b\, \Delta + 2 \kappa \,a}{2\,\kappa^2},\\
\dot p& =&\frac {p\, \Delta + 2 \kappa \,r}{2\,\kappa^2},\\
\dot r& =&\frac {r\, \Delta - 2 \kappa \,p}{2\,\kappa^2},\\
\dot c& =&c\frac {  \Delta - 2 \,Z\,\kappa}{2\,\kappa^2},\label{eqcB67}
\end{eqnarray}
with
\begin{equation}\label{Dk}
\Delta = (b^2+r^2)-(a^2+p^2),\text{ and } \kappa=a\,b-p\,r.
\end{equation}
It is obvious from these equations that $\kappa$ is a constant. Moreover, the fact that we may diagonalize the frame at $t=t_0$ implies that it remains diagonal during all of its evolution: thus $r(t)=p(t)=0$.\\

By considering the combination $X=a+i\,b$ the main equation to solve is :
\begin{equation}
  \dot X=-X  (X^2 - 4\,i\, \kappa )/(2\,\kappa^2);
  \end{equation}
after which we may finally obtain $c$ thanks to (\ref{eqcB67}).  Solving both equations yields:
\begin{eqnarray}
X =(a+i\,b) &=&2\,X_0\,\,e^{i\,t/\kappa}\,\left( i\,\kappa/(X^2_0\,e^{2\,i\,t/\kappa}-\bar X^2_0 \,e^{-2\,i\,t/\kappa})\right)^{1/2}\quad ,\\
c^2 &=&c^2_0 \,e^{-2\,Z\,t/\kappa }/( X(t)\,\bar X(t))\quad .
\end{eqnarray}
Moreover the remaining transformation with  $\Lambda^{-1}=\text{diag} (\rho,\rho,1)$ allows us to fix $\kappa=1$ . After a final translation of the variable $t$, and if necessary a flip of sign we obtain
\begin{eqnarray}
a=1/b &=&\sqrt{\cot (t)}\quad ,\\
c &=&c _0 \,\frac{e^{Z\,t }}{\sqrt{{\sin(2\,t)}}}\quad .
\end{eqnarray}
Similar expressions can be obtained for Bianchi type VII$_0$, but they involve real exponentials of time.\\

\subparagraph{Bianchi class B} In this case , condition (\ref{R1I}) is equivalent to
\begin{equation}\label{R1IB}
I_\beta n^{\alpha\beta}=\underset{\dot\strut}{\epsilon}^{\alpha\mu\nu}I_\mu\,a_\nu\qquad( \underset{\dot\strut}{\epsilon}^{123}=1).
\end{equation}
For algebras of Bianchi types IV,  VI$_h$ (with $h\neq 0,1$) and VII$_h$ (with $h\neq 0$) we found that the only solution to the equations (\ref{R1IB}) is $I_\alpha=\lambda\, a_\alpha$, but  for type
III $=$ VI$_1$, we obtain a two-parameter solution. This is a reflection of the fact that the derived algebra now is of rank one instead of two.\\

Equations (\ref{sym2}) read
\begin{equation}\label{2Isym}
\lambda_i\,\rho_j-\lambda_j\,\rho_i=-2\,\alpha^k\epsilon_{kij}\text{ with }\alpha^k=a_\mu\theta^{\mu\,k}.
\end{equation}
As we require a non singular co-frame, they are only compatible with an algebra of Bianchi type III. By derivation, using (\ref{vsym}) we obtain the system of equations:
\begin{eqnarray} \label{2Isym2}
\dot \rho_i&=&(\frac 12 n^{\mu\nu}g_{\mu\nu}+\theta \,\lambda^k\rho_k)\rho_i +2\,\theta\epsilon_i^{\ jk}\alpha_j\rho_k ,\nonumber\\
\dot \alpha_i&=&\frac 12 n^{\mu\nu}g_{\mu\nu}\,\alpha_i+\theta\,\lambda^k\,\alpha_k\,\rho_i.
\end{eqnarray}
To check the consistency of (\ref{2Isym}) with (\ref{Ifl}), let us  rewrite them as relations between vectors of a three dimensional Euclidean space:
\begin{eqnarray}
2\,\vec \alpha&=&-\vec \lambda\times\vec \rho ,\label{2alr}\\
\dot{\vec \rho}&=&\frac n 2\vec \rho +\theta\,(\vec \lambda \cdot \vec \rho)\,\vec \rho + 2\, \theta\,\vec \alpha\times \vec \rho,\label{vrdot}\\
\dot{\vec \alpha}&=&\frac n 2\vec \alpha +\theta\,(\vec \lambda \cdot \vec \alpha)\,\vec \rho .\label{vadot}
\end{eqnarray}
Taking the derivative of (\ref{2alr}) and inserting in it (\ref{vrdot}, \ref{vadot}), we obtain
\begin{eqnarray*}
2\,\dot{\vec\alpha}+\vec\lambda\times\dot{\vec\rho}&=&
n\,  \vec \alpha+2\,\theta\,(\vec\lambda\cdot\vec\alpha)\,\vec\rho
+\frac n 2 \vec \lambda\times\vec\rho + \,\theta\,(\vec\lambda\cdot\vec\rho)\,\vec \lambda\times\vec\rho+2\,\theta\,\vec\lambda\times(\vec\alpha\times\vec\rho), \\
&=&\left(\frac n 2\,\vec\alpha+ \theta\,(\vec\lambda\cdot\vec\rho)\right)(2\,\vec\alpha+\vec \lambda\times\vec\rho).
\end{eqnarray*}
Thus if $2\,\vec \alpha+\vec \lambda\times\vec \rho=0$ at a period of time, it always vanishes. This insures that the solution obeying this condition constitutes a self-dual solution.\\

It was shown in \cite{Lorenz:1983,Lorenz:1989} that there doesn't exist any diagonal self-dual metrics with a Bianchi type III symmetry group. We provide in the following the most general solution of type III, without any {\it a priori} (non geometrical) assumption on the co-frame, from which it is easy to see that diagonal metrics don't exist but non diagonal ones do. As previously, assuming a non singular frame, by using rotations and linear transformations we may put $\lambda=(1,0,0)$ and $I_\alpha=(1,1,0)$. Then (\ref{sym2}) implies that the  frame must be of the form
\begin{equation}\label{gencf3}
(\theta^\alpha_i)=
\left(
\begin{array}{ccc}
a  & 2\,c-b  &-2\,v-s   \\
r  & b  &  s \\
 0 & v  & c
\end{array}
\right).
\end{equation}
This  frame can still be transformed, without altering the canonical structure constants and constants of motion, by acting on the right with the $GL(3,\mbR)$ matrix
\begin{equation}  \label{pbB}
\Lambda^{-1}=\left(
\begin{array}{ccc}
  1- \mu& \mu  & \nu  \\
\mu  & 1-\mu  &  -\nu\\
0  &0  &1
\end{array}
\right)
\end{equation}
and on the left with a rotation that leaves $\gl$ invariant
\be
{ O} =
\left(
\begin{array}{ccc}
  1& 0  & 0  \\
0  & \cos(\phi)  &  \sin (\phi)\\
0  & -\sin (\phi)  & \cos (\phi)
\end{array}
\right).
\ee
The self-duality equations imply that $v(t)=V  c(t)$. Thus we may assume that $v(t)=0$ after, if necessary, a rotation of angle $\phi$ such that $\tan (\phi)=V$. Moreover we then see that $s(t)= S\,c(t)$ and that by choosing an appropriate value of $\nu$ we may also assume $s(t)=0$. The remaining parameter $\mu$ can by fixed by requiring $b(0)=0$.\\

So without loss of generality, the frame we have to consider reduces to:
\be \label{IIImet}
\theta_k^\ga=
\left(
\begin{array}{ccc}
a   & 2\, c -b    & 0  \\
  r &b   &0  \\
 0&  0&  c
\end{array}
\right).
\ee
In order to pursue the integration it will be useful to redefine the unknown functions as follows (using the fact that $c$ cannot vanish as we only consider non singular frames):
\begin{equation}
a(t) = \alpha(t)\, c(t),\quad b(t) = \beta(t)c(t),\quad    r(t) = \rho(t)\, c(t).
\end{equation}
Then the self-duality equations lead to
\begin{eqnarray}
\dot\alpha&=&\frac{2+\alpha^2+\alpha\,\rho}{\left(2\,\rho-(\alpha+\rho)\,\beta\right)c^2 } \ ,\\
\dot\beta&=&\frac{\rho- \alpha }{\left(2\,\rho-(\alpha+\rho)\,\beta\right)c^2 }\  ,\label{teq}\\
\dot\rho&=&\frac{2+\rho^2+\alpha\,\rho}{\left(2\,\rho-(\alpha+\rho)\,\beta\right)c^2 }\  ,\\
\dot c&=&\frac{\rho^2-\alpha^2+4\,\beta-4}{2\,c\,\left( 2\,\rho-(\alpha+\rho)\,\beta\right)^2 }\  .
\end{eqnarray}
They imply
\begin{equation}
\frac{d(\rho-\alpha)}{d\beta}=(\rho+\alpha)\text{ and }
\frac{d(\rho+\alpha)}{d\beta}=\frac{4+(\rho+\alpha)^2}{\rho-\alpha}.
\end{equation}
These equations are easy to solve and lead to
\begin{eqnarray*}
\alpha&=&\sinh\left(\frac{\beta-\beta_*}K\right)-K\,\cosh \left(\frac{\beta-\beta_*}K\right)\  ,\\
\rho&=&\sinh\left(\frac{\beta-\beta_*}K\right)+K\,\cosh \left(\frac{\beta-\beta_*}K\right)\  .\\
\end{eqnarray*}
Then we may express, by a quadrature, the function $c$ as a function of $\beta$ using the equation obtained from the ratio of $\dot \beta $ and $\dot c $:
\be c^2 =c^2_0\frac{\cosh \left(\frac{\beta-\beta_*}K\right)}{(\beta-1)\,\sinh \left(\frac{\beta-\beta_*}K\right)-K\,\cosh \left(\frac{\beta-\beta_*}K\right)}
\ee
   and finally express everything as functions of $t$ by integrating equation (\ref{teq}) which leads to the surprisingly simple result (after a flip of the sign of $t$):
\begin{equation}
\beta = {K \over c_0^2} \,t \  .
\end{equation}
The solution we discussed is generic. It depends on two arbitrary constants, namely $c_0$, $K$ and $\beta_*$, note that we have used a time translation to set $\beta = 0$ at time $t=0$.  Note that we assumed that $a$ is always different from $r$.  If for a value of $t$ we set $a=r$, then they become identical for all times. In this special case, the derivative of $\beta$ vanishes and we obtain that $b(t)=  B\,c(t)$  implying the remaining equations:
\bea \label{sdeq}
\frac{\dot c}{c} & = & \frac{R}{a^2},\nonumber \\
\frac{\dot a}{a} & = & -2\frac{R}{c^2}-\frac{R}{a^2}.
\eea
with $R= \frac{1}{2(1-B)}$.\\

To solve the equations (\ref{sdeq}) we use an auxiliary  function $y(t)=\log |a(t)\,c(t)|$ and obtain, after elementary operations, the first integral:
\be \label{eqcstrc}
 \dot y^2+4\,R^2 e^{-2\,y}=L^2,
\ee
from which we deduce the expression of the product $a(t)\,c(t)$ and
\begin{eqnarray}
c(t)^2&=&\frac{2\,R}L\coth[L\,t],\\
a(t)^2&=&\frac RL\sinh[2\,L\,t].
\end{eqnarray}
Let us notice that $R\,t$ has to be positive for the solution to be real. When $t=0$ we encounter a curvature singularity.

\section{Self-dual Connections }\label{sdcon}

In the conventions we have used, we naturally find solutions whose curvature is self-dual while the connection is not.  However, as we will show below, we may always make a local $SO(4)$ rotation to make the connection self-dual.  In fact, the vanishing of the anti self-dual part of the connection can be related to the integrability condition, thereby ensuring the existence of a pure self dual-connection. We shall illustrate this point by evaluating, in general, the gauge transformations which map our canonical co-frame to one whose associated connection is self-dual.

As is well known, the two (families of) parallelisms on the three-sphere $S^3$ provide a way to decompose any $SO(4)$ rotation into the product of self-dual and anti self-dual $SO(3)$ rotations. If $u$, $x$, $y$ and $z$ are the coordinates of a point of $S^3$ : $u^2+x^2+y^2+z^2=1$, an anti self-dual $SO(3)$ element can be written as:
\begin{equation}\label{selfrot}
O_{ad}:=
\left(
\begin{array}{cccc}
 u & -x & -y & -z \\
 x & u & z & -y \\
 y & -z & u & x \\
 z & y & -x & u
\end{array}
\right) .
\end{equation}
The gauge transformation generated by such a rotation leaves invariant the self-dual part of the connection but transforms the anti self-dual one. Accordingly, when the anti self-dual part of the connection, given by the constants $\bar I_{i\ga}$, is non-zero, such a gauge transformation may set it to zero thanks to a solution of the equation:
\begin{equation}\label{dLwL}
dO=O \ \underset a \omega.
\end{equation}
More explicitly we obtain (with $\bar{\bf  I}_k:=\bar I_{k\ga}\,\sigma^\ga$ and $x^i:=(x,y,z)$)
\begin{equation}\label{gt}
du=\frac 12 x^k\,\bar {\bf I}_k, \qquad dx^i=\frac 12 \left(- u\,\bar  {\bf I}_i+
\epsilon^{ijk} x_j\,\bar  {\bf I}_k\right).
\end{equation}
As a first immediate consequence we see that this gauge transformation has to be  $t$ independent.  Moreover we also obtain from Eq. (\ref{mainalg}):
\begin{equation}\label{dI}
d\bar {\bf I}_i=\frac 12 \bar I_{i\ga}c^\ga_{\ \beta\gamma}\,\sigma^\beta\wedge \sigma^\gamma=-\frac 12 \epsilon_i^{jk}\bar  {\bf I}_{j }\wedge \bar  {\bf I}_{k },
\end{equation}
which indeed ensures the integrability of the equations (\ref{gt}). In the case of a rank one mapping, $\bar I_{k\alpha}=\lambda_k\,I_\alpha$ (normalized such that $\sum_k\lambda_k^2=1$), this equation implies there exists a function $\chi$ such that $d\chi=I_\alpha\,\sigma^\alpha$.  Correspondingly,
the gauge transformation (\ref{selfrot}) that leads to a frame which has vanishing anti self-dual connection is given by
\begin{equation}\label{sdfr1}
u=\sin(\chi/2), \qquad x^k=\lambda^k\,\cos(\chi/2).
\end{equation}

In the case of rank three, the only possibility is Bianchi type IX.  As discussed above it can be taken diagonal, with the matrix $(\bar I_{i\alpha})=-\delta_{i \alpha}$.  Moreover, it is well known that the components of the invariant one-forms also define parallel vector fields. Using the identities $u\, du + x\,dx+y\,dy+z\,dz=0$ and $u^2+x^2+y^2+z^2=1$, it is easy to check that the invariant one-forms may be expressed as~:
\begin{eqnarray}
\sigma^1&=&-2 (x\, du -u\,dx+z\,dy-y\,dz),\\
 \sigma^2&=&-2 (y\, du -z\,dx-u\,dy+x\,dz),\\
  \sigma^3&=&-2 (z\, du +y\,dx-x\,dy-u\,dz),
\end{eqnarray}
which satisfy the invariance relations: $d\sigma^i=\epsilon^i_{j\,k}\sigma^j\wedge\sigma^k$ as well as (\ref{gt}). In the case where the matrix $(\bar I_{i\alpha})$ is not diagonal, it must be (in order to satisfy to equations (\ref{vecB9})) of the form $(\bar I_{i\alpha})=-(O_{i\alpha})$ with $O$ an $SO(3)$ rotation matrix.  It is elementary to verify that the equations (\ref{gt}) are solved by similar functions involving variables $u$ and $(x'^k)=(x',y',z')$ related to $(x^k)=(x,y,z)$ by the rotation $O$:
 \begin{equation}
 x'k=-I_{kl}x^l
\end{equation}
as it must be.

\section{Conclusion}

In this note we have achieved a complete classification of all the self-dual Euclidean spaces admitting a $G3$ simply transitive homogeneity group. For each Bianchi group, there are as many classes of solutions as homomorphisms $\bar I:\frak{g}_3\to \frak{so}(3)$. For each of these homomorphisms, labeled by its rank, the self-duality constraint is  described by a specific set  of equations -- like Lagrange or Darboux--Halphen in Bianchi IX, leading to distinct solutions such as Eguchi--Hanson or Taub--NUT respectively. These equations have non-everywhere vanishing solutions for all class  A Bianchi groups (I, II, VI$_{0}$, VII$_{0}$, VIII and IX), whereas in class   B solutions exist only for Bianchi III. The latter solutions  (Eqs. (\ref{IIImet1}),  (\ref{IIImet2})) are \emph{all} rank-$1$ and non-diagonal, contrary to the lore  that it seamed  ``very unlikely to construct non-diagonal self-dual solutions of Bianchi types I--IX''   \cite{Lorenz:1983,Lorenz:1989}. This expectation was empty for Bianchi A because in the present formulation, the diagonal ansatz is the most general, but not for Bianchi B.\\

It is fair to stress that the above conclusions have been reached by adopting a metric representation, Eq.  (\ref{genmet}), such that the foliation of $\mathcal{M}$ as $\mathbb{R}\times \Sigma_3$ is manifest and adapted to the splitting of the group $SO(4)$ into self-dual and anti-self-dual factors. Other choices for the metric may exist, where the distinction between the various classes of self-dual solutions is less sharp. This happens e.g. for Bianchi IX, in the Gibbons--Hawking  representation \cite{Gibbons:1979xm}, where Eguchi--Hanson (rank-$0$) and Taub--NUT (rank-$1$) are indeed, to some extent, unified. This is possible, however, at the price of abandoning
the $G3$-invariant frames that we use. Here, this choice has been instrumental, not only for providing the classification in terms of the rank of the  $\frak{g}_3\to \frak{so}(3)$ homomorphism, but also for scanning all possible solutions.\\

Our analysis can easily be adapted to the ultra-hyperbolic case (spacetime signature $(-,-,+,+)$), where the algebra of Bianchi type VIII will play a role analogous to the one of Bianchi IX here. It should however be emphasized that we will have to distinguish, in that case, among the inequivalent one-dimensional subalgebras of ${\frak so}(2,1)$. Other extensions concern the addition of a cosmological constant (Weyl self-duality), or the implementation of the method to higher-dimensional set-ups admitting self-duality, as e.g. in seven dimensions. Last but not least, the physical analysis of the new Bianchi III solutions remains to be completed. The large number of moduli in (\ref{IIImet1}), (\ref{IIImet2}) make such an analysis quite involved.\\

It would be interesting to discover if the techniques and classification presented in this paper can be generalized to cases where we have a more general setup than vacuum Einstein equations.  For example, the evolution of certain Bianchi spaces in the presence of a cosmological constant, as well as various matter such as a perfect fluid and/or scalar, spinor, and electro-magnetic fields has been studied recently in \cite{Barrow:1999qq,Saha:2007dr,Saha:2004vu}.  The authors have derived exact solutions in certain cases, some of which are singularity free, and a complete classification of solutions would be appreciated.  In the case of Einstein's equations with a cosmological constant, it is well known that requiring self-duality of the Weyl tensor is sufficient to satisfy the equations of motion. However in this case our technics have to be abandoned as we are generically confronted  with a system of second order differential equations.  

We note that most of the Bianchi classes have solutions which necessarily have singularities.  It would be interesting to discover if these singularities could be cured by the introduction of either a cosmological constant and/or additional matter fields.  Examples have already been discussed in \cite{Barrow:1999qq,Saha:2007dr,Saha:2004vu} as well as \cite{Spokoiny:1982pa,Fay:2000zm}.

\section*{Acknowledgements}

The authors thank P. Bieliavsky for stimulating discussions and a referee for having indicated us the paper \cite{Barrow:1999qq}.  Fran\c cois Bourliot, J. Estes and M. Petropoulos would like to thank the Service de Physique
Th\'eorique de l'Universit\'e Libre de Bruxelles as well as the Service de M\'ecanique et Gravitation de l'Universit\'e de Mons--Hainaut for kind hospitality. Philippe Spindel thanks the IHES where the
present collaboration was initiated and the CPHT of Ecole Polytechnique, and acknowledges financial support from
IISN-Belgium (convention 4.4511.06). John Estes acknowledges financial support from the Groupement d'Int\'er\^et Scientifique P2I. This research was partially supported by the French Agence Nationale pour la Recherche, contract  09-BLAN-NT09-573739 and by the CNRS-FNRS-CGRI-2009.

\appendix

\section*{Appendix: G3 self-dual metrics}

We summarize here all known (real) self-dual $G3$-homogeneous gravitational instantons, following the classification pattern we have developed in terms of homomorphisms $\frak{g}_3\to \frak{so}(3)$. These can be of rank $0,1$ or $3$. Whenever rank-$0$ and rank-$1$ solutions coexist (\emph{i.e.} for Bianchi  I, II, VI$_0$ and VII$_0$), they are both captured by a single expression with a two-valued parameter $\epsilon=0,1$. Bianchi IX is the only case which possess rank-0 and rank-3 solutions. Bianchi III is the most peculiar: contrary to the other classes, it requires a non-diagonal co-frame (\ref{gencf}) (and consequently a non-diagonal metric (\ref{genmet})) and allows exclusively rank-$1$ solutions. These solutions are new.\\

In the following presentation, integration constants that can be reabsorbed by coordinate redefinitions have been discarded. Consequently, all remaining parameters are \emph{genuine moduli} of the solutions. \\

 \paragraph{Bianchi I}
The solution reads
\be
ds^2=  e^{-2\,\epsilon\,t}\,dt^2+e^{-2\,\epsilon\,t}\,(\sigma^1)^2+(\sigma^2)^2+(\sigma^3)^2
\ee
and describes  flat geometry for both rank-$\epsilon$ solutions ($\epsilon=0,1$). Notice, however, that the rank-$0$ case appears naturally in Cartesian coordinates, whereas  rank-$1$ emerges in a kind of  mixed Cartesian/polar,  Euclidean--Rindler-like coordinates.

\paragraph{ Bianchi II}

The rank-$\epsilon$ solutions are given by
\be
ds^2= \frac{t \, e^{2 \, \epsilon\,t /b_0}}{b_0^2\,c_0^2} dt^2+\frac{1}{t} (\sigma^1)^2+  \frac{t}{b_0^2} (\sigma^2)^2+ \frac{t \, e^{2 \, \epsilon\,t /b_0}}{c_0^2}(\sigma^3)^2.
\ee
The Kretschmann scalar is given by
\be
{\cal K} = R_{MNPQ} R^{MNPQ} = 8\,b_0^2\,c_0^4\,e^{-4\,t\,\epsilon/b_0} \frac{(\epsilon\,t)^2+3\,b_0\,\epsilon\,t+3\,b_0^2}{t^6}
\ee
from which we can see that at $t=0$ the metric has a curvature singularity.

 \paragraph{Bianchi III}

The Bianchi-III self-dual metrics are all rank-one. They are captured by two expressions, depending on whether $a\neq r$ (Eq. (\ref{IIImet1})), or  $a= r$  (Eq. (\ref{IIImet2})),  in the general co-frame (\ref{gencf}) after a convenient rescaling of $t$ by $t \rightarrow c_0^2 t$ we obtain:
\begin{eqnarray}
ds^2&=&
\frac{F(t)}{32\,c_0^4\,\cosh^4 \tmts} dt^2 +
\frac{ \big(K\,t-1\big) \tanh\tmts - K}{c_0^2} (\sigma^3)^2 \nonumber\\
&&
+ \frac{g_1(t)}{F(t)} (\sigma^1)^2 + \frac{g_2(t)}{F(t)} (\sigma^2)^2 + 2\,\frac{g_3(t)}{F(t)} \, \sigma^1\,\sigma^2
,\label{IIImet1}
\end{eqnarray}
where we have introduced the functions $F(t)$, $g_1(t)$, $g_2(t)$ and $g_3(t)$ given by
\bea
F(t) &=& 8 \, c_0^2 \, \cosh \tmts \bigg[\big(K\,t - 1\big) \sinh \tmts - K\,\cosh \tmts \bigg], \nonumber \\
g_1(t) &=& (K^2 + 1) \cosh\twotmts + 2\,K\,\sinh\twotmts + 2 K^2\,t^2 + K^2-1, \nonumber \\
g_2(t) &=& (K^2 + 1) \cosh \twotmts - 2\,K\,\sinh \twotmts \nonumber\\ && + 2 K^2\,t^2 - 8\, K\,t + K^2+7, \nonumber\\
g_3(t) &=& (K^2-1)\cosh \twotmts + 2\,K^2\,t^2 - 4\,K\,t + K^2+1.
\eea
Denoting the zeroes of $F(t)$ as $t_i$, we find that the Kretschmann scalar diverges as $(t-t_i)^6$ as $t \rightarrow t_i$, indicating a curvature singularity at each of the two zeros of $F(t)$.
The second solution with $a= r$ is given by
\begin{eqnarray}
ds^2&=&\frac  {L\,(B-1)}4 \frac {\tanh(t)}{\cosh^2(t)}\,dt^2 + L\,(B-1)\,\tanh(t)\, \left(\sigma^3\right)^2\nonumber\\
 &&+
 \frac{L}{4\,(B-1)}\left\{ \left[2\,B^2\,\text{csch}(2\,t)+\tanh(t)\right]\left(\sigma^1\right)^2+ \right.\nonumber\\
  &&- 2 \, [\tanh(t)-2\,B\,(B-2)\text{csch}(2\,t) ]\,  \sigma^1\,\sigma^2  \nonumber\\
&&\left. \left [\text{csch}(2\,t)(\cosh(2\,t)+2\,(B-2)^2-1\right]\left(\sigma^2\right)^2\right\}.\label{IIImet2}
\end{eqnarray}
In this case, the Kretschmann scalar is given by
\be
{\cal K} = \frac{384 \, \coth^6(t)}{(B-1)^2 L^2}
\ee
and we find at $t=0$ the metric has a curvature singularity.  The general analysis of these geometries is involved and provides interesting features, which deserve a separate study.

 \paragraph{Bianchi VI$_{0}$}

The general metric for rank-$\epsilon$ solutions reads:
\be
ds^2=\sin(2\,t) \,\frac{e^{2\,\epsilon\,Z\,t }}{c^2_0} \left( dt^2+ \left(\sigma^3\right)^2
\right)+\tan (t)\left(\sigma^1\right)^2+\cot (t)\left(\sigma^2\right)^2,
\ee
where  $t\in [0, {\pi}/{2}]$. A curvature singularity appears at the boundaries $t=0$ and $t=\pi/2$.
The Kretschmann scalar is given by
\be
{\cal K} = 16 \, c_0^4 \, e^{-4 \, t \, \epsilon\,Z} \frac{6\,\epsilon\,Z\,\sin(4\,t)+[9+(\epsilon\,Z)^2]\cos(4\,t)+(\epsilon\,Z)^2+15}{\sin^6(2\,t)},
\ee
and we see that at $t=0, \pi/2$ the metric has a curvature singularity.

\paragraph{Bianchi VII$_{0}$}

The general metric for rank-$\epsilon$ solutions reads:
\be
ds^2=\sinh(2\,t) \, \frac{e^{2\,\epsilon\,Z\,t }}{c^2_0} \left( dt^2+ \left(\sigma^3\right)^2
\right)+\tanh (t)\left(\sigma^1\right)^2+\coth (t)\left(\sigma^2\right)^2
\ee
The Kretschmann scalar is given by
\be
{\cal K} = 16 \, c_0^4 \, e^{-4 \, t \, \epsilon\,Z} \frac{6\,\epsilon\,Z\,\sinh(4\,t)+[9-(\epsilon\,Z)^2]\cosh(4\,t)-(\epsilon\,Z)^2+15}{\sinh^6(2\,t)}
\ee
and we again find a curvature singularity at $t=0$.
In the rank-$1$ case, we may take the large-$t$ limit of the above metric to obtain
\be
ds^2=c^2_0 \, {e^{2(1+Z)t }} \left( dt^2+ \left(\sigma^3\right)^2
\right)+\left(\sigma^1\right)^2+\left(\sigma^2\right)^2
\ee
which also solves the self-duality equations but has a vanishing Kretschmann scalar and is actually flat.

\paragraph{Bianchi VIII}

{There is only the rank-$0$ solution whose simplest writing is:
\be
ds^2=\frac{P^{-1/2}}{4}dx^2+P^{1/2}\left(\frac{(\sigma^1)^2}{x_1-x}+\frac{(\sigma^2)^2}{x_2-x}+\frac{(\sigma^3)^2}x \right)\label{B8alg}
\ee
with $P=(x_1-x)( x_2-x)x$ where $x_1$, $x_2$ are positive and $0\leq x\leq \min\{x_1,x_2\}$. The Krestschmann scalar blows up as $x^{-3}(x-x_1)^{-3}(x-x_2)^{-3}$, indicating that the singularity at the boundary of the domain of definition of $x$ is a true one. Let us assume $x_1= \alpha^2>x_2= \beta^2$. We may re-express the metric in the gauge $N=\theta$, using Jacobi elliptic functions of module $\beta/\alpha$. Making the substitution  $x=\beta^2\,\text{sn}^2(\alpha\,t)$ we obtain:
\begin{eqnarray}
ds^2&=&\alpha\,\beta^2  \text{sn}(\alpha\,t)\,\text{cn}(\alpha\,t)\,\text{dn}(\alpha\,t) dt^2  \\
&&+\frac{\beta^2\,\text{sn}(\alpha\,t)\,\text{cn}(\alpha\,t)}{\alpha\,\text{dn}(\alpha\,t)}(\sigma^1)^2+\frac{\alpha\,\text{sn}(\alpha\,t)\,\text{dn}(\alpha\,t)}{ \text{cn}(\alpha\,t)}{(\sigma^2)^2}+\frac{\alpha\,\text{cn}(\alpha\,t)\,\text{dn}(\alpha\,t)}{ \text{sn}(\alpha\,t)}{(\sigma^3)^2}.\nonumber
\end{eqnarray}}
The $t$ variable belongs to the interval $[0,2\,K( \beta/\alpha)/\alpha]$,  $K(k)$ being the
complete elliptic integral of the first kind of module $k$.

\paragraph{Bianchi IX}
We have solutions in case of rank-$0$ (\textit{Lagrange} system of equations) and of rank-$3$ (\textit{Darboux-Halphen} system of equations).

 {The general rank-$0$ solution \footnote{ The metric can be chosen diagonal; the off-diagonal entries are consistently set to zero in this class.}
was found by Belinsky \emph{et al } in  \cite{Belinsky:1978ue} as a strict
$SU(2)$-symmetric generalization of the Eguchi--Hanson gravitational instanton. The latter has enhanced $SU(2)\times U(1)$ isometry and is a solution of the Lagrange system. An algebraic expression of the solution, analogous in  form to Eq. (\ref{B8alg}),  is
\be
ds^2=\frac{P^{-1/2}}{4}dx^2+P^{1/2}\left(\frac{(\sigma^1)^2}{x-x_1}+\frac{(\sigma^2)^2}{x-x_2}+\frac{(\sigma^3)^2}x \right)
\ee
with $P=(x-x_1)(x-x_2)x$ where $x_1$, $x_2$ are negative. The general strict $SU(2)$-symmetric solutions have curvature singularities.  To obtain the Eguchi--Hanson gravitational instanton, one may take $x_1 = x_2 =x_0$ and by a translation in $x$ set $x_3 = 0$.\\
Let us assume $x_1=-\alpha^2<x_2=-\beta^2$. In the gauge $N=\theta$, the metric can still be expressed in terms of standard Jacobi elliptic functions but now of module $\sqrt{\alpha^2-\beta^2}/\alpha$; using $x=\beta^2 \text{sn}^2(\alpha\,t)/\text{cn}^2(\alpha\,t)$ with $t\in[0,K(\sqrt{\alpha^2-\beta^2}/\alpha)/\alpha]$ we obtain:
\begin{eqnarray}
ds^2&=&\alpha\,\beta^2 \frac{\text{sn}(\alpha\,t)\,\text{dn}(\alpha\,t)}{\text{cn}^3(\alpha\,t)}dt^2
\\&&+\frac{\beta^2\,\text{sn}(\alpha\,t)}{\alpha\,\text{cn}(\alpha\,t)\,\text{dn}(\alpha\,t)}(\sigma^1)^2+\frac{\alpha\,\text{sn}(\alpha\,t)\,\text{dn}(\alpha\,t)}{ \text{cn}(\alpha\,t)}{(\sigma^2)^2}+\frac{\alpha\,\nonumber\text{dn}(\alpha\,t)}{ \text{sn}(\alpha\,t)\,\text{cn}(\alpha\,t)}{(\sigma^3)^2}.
\end{eqnarray}}

The issue of the rank-$3$ solution is more subtle. It is a generalization of Taub--NUT metric, with strict $SU(2)$ isometry, and solves the Darboux--Halphen system. As explained in \cite{ Takhtajan:1992qb},  it requires the use of modular forms. For simplicity we trade $a,b,c$ for $\Omega_\alpha,\,  \alpha=1,2,3$, where $\Omega_1 = \frac{1}{b\,c}$, $\Omega_2 = \frac{1}{a\,c}$, and $\Omega_3 = \frac{1}{a\,b}$, and introduce a triplet of weight-two modular forms of $\Gamma(2)\subset PSL(2,\mathbb{Z})$:
\be
E_1=\frac{{d\lambda}/{dz}}{\lambda},\ E_2=\frac{{d\lambda}/{dz}}{\lambda-1},\ E_3=\frac{{d\lambda}/{dz}}{\lambda(\lambda-1)},
\ee
where $\lambda$ is solution of Schwartz' equation
\be
\frac{d^3 \lambda/dz^3}{d\lambda/dz}-\frac{3}{2}\left(\frac{d^2\lambda/dz^2}{d\lambda/dz}\right)^2=-\frac{1}{2}\left(\frac{1}{\lambda^2}+\frac{1}{(\lambda-1)^2}-\frac{1}{\lambda(\lambda-1)}\right)\left(\frac{d\lambda}{dz}\right)^2.
\ee
Any real solution of the Darboux--Halphen system reads:
\be
\Omega_\alpha(t)=-\frac{1}{2}\frac{d}{dt}\log E_\alpha(it),\ \alpha=1,2,3,
\ee
and generates a Bianchi-IX, rank-$1$ gravitational instanton (Eqs. (\ref{genmet}) and (\ref{gencf})). A generic curvature singularity is present in these geometries. This singularity can be pushed at infinity for one specific choice of $\lambda(z)$:  $\lambda={\theta_2^4}/{\theta_3^4}$, where $\theta_{2,3}$ are the standard Jacobi theta functions\footnote{For concreteness ($q=\exp{2i\pi z}$)
$$
\theta_{2}(z)=  2q^{\frac{1}{8}}\prod_{n=1}^{\infty} (1-q^{2n})(1+q^{n}),\quad
\theta_{3}(z)  =  \prod_{n=1}^{\infty} (1-q^{n})(1+q^{n-1/2})^2.
$$}.
This case corresponds to the Atiyah--Hitchin solution  \cite{sol1985,sol19852}.  Similar to the rank-$0$ solution, in the case that two of the metric factors are equal the solutions has enhanced $SU(2)\times U(1)$ isometry and takes the form of the well-known Taub-NUT metric
\bea
ds^2 = \frac{r^2}{4(1+k\,r^2)^2} \big( (\sigma^1)^2 + (\sigma^2)^2 \big) + \frac{r^2}{4} (\sigma^3)^2 + \frac{1}{(1+k\,r^2)^4}dr^2.
\eea

\bibliographystyle{plain}
\bibliography{MaBiblio}{}

\end{document}